\documentclass[a4paper,11pt]{article}
\pdfoutput=1 

\usepackage{jcappub} 
 \usepackage{indentfirst}
\usepackage{appendix}

\title{Revisit of constraints on dark energy with Hubble parameter measurements including future redshift drift observations}

\author[a]{Yan Liu,}
\author[a]{Rui-Yun Guo,}
\author[a]{Jing-Fei Zhang,}
\author[a,b,1]{Xin Zhang\note{Corresponding author.}}
\affiliation[a]{Department of Physics, College of Sciences, Northeastern University, \\Shenyang
110819, China}
\affiliation[b]{Center for High Energy Physics, Peking University, \\Beijing 100080, China}

\emailAdd{liuyanld@163.com, guoruiyun110@163.com, jfzhang@mail.neu.edu.cn, zhangxin@mail.neu.edu.cn}
\abstract{We investigate whether the current Hubble parameter $H(z)$ measurements could help improve the constraints on dark energy on the basis of the mainstream cosmological probes including the type Ia supernovae (SN) observation, the cosmic microwave background anisotropies (CMB) observation, and the baryon acoustic oscillations (BAO) observation. For the current $H(z)$ data, we use 30 data points measured by using a differential age method. Furthermore, we also consider the future $H(z)$ measurements based on the Sandage-Loeb (SL) test by means of the E-ELT in construction, and thus we also use 30 simulated $H(z)$ data according to a 10-year SL test observation. In this work, we choose four typical dark energy models as examples, i.e., the $\Lambda$CDM model, the $w$CDM model, the $\alpha$DE model, and the GCG model, to complete the analysis. We find that, when only the current $H(z)$ data are added, the constraints on these models are not improved compared to the cases using the SN+CMB+BAO data; but when further adding the 10-year SL test data, the constraint results are tremendously improved for all the four models. Therefore, we conclude that, although the current $H(z)$ measurements could not provide an evident improvement on the basis of the current mainstream cosmological probes, the future $H(z)$ measurements from the SL test would have enormous potential to change the status of the Hubble parameter measurements in constraining dark energy.}

\begin{document}
\maketitle
\flushbottom

\section{Introduction}
\label{sec1}
The accelerating expansion of the universe was discovered by two observation teams through the observations of type Ia supernovae~\cite{Riess:1998cb,Perlmutter:1998np} and was subsequently confirmed by various other astronomical observations~\cite{ Spergel:2003cb,Bennett:2003bz,Tegmark:2003ud,Abazajian:2004aja}. In order to explain the phenomenon of the cosmic acceleration, a new energy component in the universe, named dark energy, is proposed, which is assumed to be a source to produce negative pressure and can thus drive the cosmic acceleration~\cite{ Sahni:1999gb,Padmanabhan:2002ji,Peebles:2002gy,Copeland:2006wr,Sahni:2006pa,Frieman:2008sn,Li:2011sd,Bamba:2012cp,
Weinberg:2012es,Mortonson:2013zfa}. The current observations reveal that dark energy contributes about $68\%$ of the total energy in the universe. However, we actually know little about the nature of dark energy. To investigate the nature of dark energy, numerous dark energy models have been proposed.

The preferred candidate of dark energy is the so-called cosmological constant $\Lambda$, first introduced by Einstein, which is actually physically equivalent to the vacuum energy density. The cosmological constant $\Lambda$  has an equation of state (EoS) $w_\Lambda=p_\Lambda/\rho_\Lambda=-1$, due to the fact that its corresponding density is a constant. The cosmological model with $\Lambda$ and cold dark matter (CDM) is usually called the $\Lambda$CDM model, which is favored by the current cosmological observations, especially by the observation of the Planck  satellite mission~\cite{Ade:2015xua}. However, it has always been challenged by some  theoretical problems, such as the fine-tuning and coincidence problems~\cite{Weinberg,Zeldovich}. Thus, an important mission in cosmology is to probe if there is some dynamics in dark energy, which usually leads to an EoS of dark energy which deviates from $-1$ and evolves in time.

It is of great difficulty to measure the EoS $w(z)$ (with $z$ being the redshift factor of the universe) of dark energy because $w(z)$ is not a direct observable. Actually, $w(z)$ has a fairly complex relationship with the observables in cosmology. To probe the expansion history of the universe, in practice one usually tries to establish a distance--redshift relation because the luminosity distance and the angular diameter distance are the most common observables in cosmology. However, these distances link to the EoS of dark energy by an integral over $1/H(z)$ (with $H(z)$ being the Hubble parameter of the universe), and $H(z)$ is affected by dark energy through another integral over a factor involving $w(z)$.
The current mainstream mature cosmological probes measuring the cosmic distances mainly include: type Ia supernovae (SN), cosmic microwave background (CMB), and baryon acoustic oscillations (BAO). By using these cosmological probes, the EoS of dark energy has been constrained to be in a precision of less than $\sim 4\%$ (assuming $w$ is a constant)~\cite{Ade:2015xua}. The aim in the next decade is to measure the EoS of dark energy to be in a precision of less than $\sim 1\%$. Obviously, for constraining $w$, the measurements of the Hubble parameter at different redshifts are of vital importance, because there is only one integral between $H(z)$ and $w(z)$.

Although the measurements of $H(z)$ are a challenging mission in cosmology, through the great efforts of astronomers some $H(z)$ data have been accumulated.
In the past decades, roughly $50$ data points of $H(z)$ have been obtained by using two astrophysical methods, namely, the measurement of differential age of galaxies and the measurement of clustering of galaxies or quasars~\cite{Moresco:2012,Farooq:2012ju,Chuang:2012hq,Farooq:2013hq,Farooq:2013eea,Chen:2011hq,Guo:2015gpa,Gaztanaga:2009hq,Wang:2016wjr,H1,H2,H3,H4,H5,H6,Anderson:2013oza,Font-Ribera:2013wce,Delubac:2014aqe,Geng:2018pxk,Simon:2004tf,Zhang1010,Farooq1211,Farooq1212,Melia:2013hsa,Li:2014yza,Sahni:2014ooa,Chen:2013vea,Park:2018tgj,Park:2017xbl,Park:2018bwy,Park:2018fxx}. These data points trace the cosmic expansion rate up to $z\approx 2$.
Actually, there has been a proposal named Sandage-Loeb (SL) test \cite{sandage,Loeb:1998bu}, which proposes measuring the redshift drifts for the quasars in the redshift range of $2\lesssim z \lesssim5$ (this range is also known as the ``redshift desert'').\footnote{This is true for the measurements of the redshift drift with the European Extremely Large Telescope (E-ELT) by observing the Lyman-$\alpha$ absorption lines of quasars. We note here that observations at $z<1$ will likely be realized by the Square Kilometre Array (SKA) \cite{Klockner:2015rqa,Martins:2016bbi} or 21 cm experiments such as CHIME \cite{Yu:2013bia}. SKA will measure the redshift drift through the observations of neutral hydrogen (HI) emission signal of galaxies at two different epochs to a precision of one percent in the range of $0<z<1$ (for the SKA Phase 2 array) \cite{Klockner:2015rqa,Martins:2016bbi}. Thus, actually, E-ELT and SKA ideally complement each other because the E-ELT probes the deep matter era and the SKA probes the acceleration era.
However, in this work, we only focus on the redshift drift measurements by the E-ELT, because our aim is to discuss how the high-redshift $H(z)$ measurements help break the parameter degeneracies formed by the low-redshift observations.}
Through the SL test, in the future one would be capable of measuring $H(z)$ up to $z\sim 5$.
These redshift-drift data would provide an important supplement to other observational data and play a fairly significant role in geometric measurements of the expansion of the universe. Recently, there have been a host of works discussing the future observations of the redshift drifts~\cite{Corasaniti:2007bg,Balbi:2007fx,Zhang:2007zga,Liske:2008ph,Zhang:2010im,Quercellini:2010zr,Martinelli:2012vq,Li:2013oba,Zhang:2013zyn,sl5,Geng:2014hoa,Geng:2014ypa,Geng:2015hen,Guo:2015gpa,He:2016rvp,Geng:2018pxk,Lazkoz:2017fvx}, showing that the simulated $H(z)$ data at high redshifts could help to break degeneracies among parameters by providing additional accurate information about $\Omega_{\rm m} h^2$.



In this work, we wish to investigate if the current $H(z)$ data could help improve the constraints on dark energy on the basis of the mainstream cosmological probes (including SN, CMB, and BAO). Furthermore, we wish to show how the SL test (future measurements of the redshift drifts) changes the status of the Hubble parameter measurements in constraining dark energy.

We will employ some concrete dark energy models to complete such an analysis. There actually have been a lot of dark energy models in the research field of dark energy, but we of course cannot use them all to do the analysis, and we can only choose several typical models as examples to complete the analysis.

In a recent work by Yue-Yao Xu and Xin Zhang \cite{Xu:2016grp}, a comparison was made for ten typical, popular dark energy models according to their capability of fitting the current observational data. They show \cite{Xu:2016grp} that according to the capability of explaining the observations the $\Lambda$CDM model is still the best one among all the models; the generalized Chaplygin gas (GCG) model, the $w$CDM model (in which $w$ is a constant), and the $\alpha$ dark energy ($\alpha$DE) model are worse than the $\Lambda$CDM model, but still are good models compared to others; the holographic dark energy (HDE) model, the new generalized Chaplygin gas (NGCG) model, and the Chevalliear-Polarski-Linder (CPL) model can still fit the current observations well, but from the perspective of ``Occam's Razor'', they are not so good; and, the new agegraphic dark energy (NADE) model, the Dvali-Gabadadze-Porrati (DGP) model, and the Ricci dark energy (RDE) model are evidently excluded by the current observations. According to this analysis result, we decide to choose the $\Lambda$CDM model, the $w$CDM model, the $\alpha$DE model, and the GCG model as typical examples to make an analysis in the present work.

The structure of this paper is arranged as follows. In Sect.~\ref{method and data}, we present the analysis method and the observational data used in this work. In Sect.~\ref{sec:Results and Discussions}, we report the constraint results and make relevant discussions. In Sect.~\ref{sec:Conclusion}, we give the conclusion of this work.

\section{Method and data}\label{method and data}

In this work, we will constrain the four typical dark energy models (the $\Lambda$CDM model, the $w$CDM model, the $\alpha$DE model, and the GCG model) by using several combinations of observational data sets, from which we wish to show: (1) whether the current $H(z)$ data could help improve the constraints on dark energy on the basis of the mainstream cosmological probes including SN, CMB, and BAO, and (2) how the SL test changes the status of the Hubble parameter measurements in constraining dark energy. Therefore, we will use the three data combinations, i.e., SN+CMB+BAO, SN+CMB+BAO+$H(z)$, and SN+CMB+BAO+$H(z)$+SL, in this work, where $H(z)$ denotes the current $H(z)$ data and SL denotes the future $H(z)$ data from redshift-drift measurement of SL test simulated according to a 10-year observation.

In the following, we shall describe briefly these models and data considered in this work.

\subsection{A brief description of the dark energy models}

Though the four dark energy models have been discussed previously in the literature, in particular they have been uniformly constrained and compared in Ref.~\cite{Xu:2016grp}, in this paper in order to be self-contained in the contents we will still make a brief description for them. In general, for the EoS of a dark energy $w(z)$, the evolution of the Hubble parameter is expressed as $E^2(z)\equiv{H^{2}(z)}/{H^{2}_{0}}=\Omega_{\rm m}(1+z)^{3}+\Omega_{\rm r}(1+z)^{4}+(1-\Omega_{\rm m}-\Omega_{\rm r})\exp [3 \int^{z}_{0} {(1+w(z'))}/({1+z'}) dz']$, where $\Omega_{\rm m}$ and $\Omega_{\rm r}$ are the present-day fractional densities of matter and radiation, respectively, and a flat universe is assumed throughout this work.

In what follows we will directly give the expressions of $E(z)$ for the specific dark energy models.

\begin{itemize}
\item $\Lambda$CDM model: Although the cosmological constant $\Lambda$ has been suffering the severe theoretical puzzles, it can explain the various observations quite well, and thus it is still the most promising candidate for dark energy. The cosmological constant has the EoS of $w=-1$, and thus we have
\begin{equation}
E^2(z)=\Omega_{\rm{m}}(1+z)^{3}+\Omega_{\rm{r}}(1+z)^{4}+(1-\Omega_{\rm{m}}-\Omega_{\rm{r}}).
\end{equation}

\item $w$CDM model: This model is the simplest case for a dynamical dark energy because the EoS of dark energy in this model is assumed to be a constant, i.e., $w={\rm constant}$. Although it is hard to believe that a constant EoS would correspond to the real physical situation, it can describe dynamical dark energy in a simple way. In this model, we have
\begin{equation}
E^2(z)=\Omega_{\rm{m}}(1+z)^{3}+\Omega_{\rm{r}}(1+z)^{4}+(1-\Omega_{\rm{m}}-\Omega_{\rm{r}})(1+z)^{3(1+w)}.
\end{equation}

\item $\alpha$DE model: This model \cite{Dvali:2003rk} is a phenomenological extension of the DGP model, in which the Friedmann equation is modified as $3M_{\rm Pl}^2(H^2-H^\alpha/r_{\rm c}^{2-\alpha})=\rho_{\rm m}(1+z)^3+\rho_{\rm r}(1+z)^4$, where $\alpha$ is a phenomenological parameter, $M_{\rm Pl}$ is the reduced Planck mass, $\rho_{\rm m}$ and $\rho_{\rm r}$ are the present-day densities of matter and radiation, respectively, and $r_{\rm c}=(1-\Omega_{\rm m}-\Omega_{\rm r})^{1/(\alpha-2)}H_0^{-1}$. In this model, $E(z)$ is derived by solving the following equation
\begin{equation}
E^2(z)=\Omega_{\rm{m}}(1+z)^{3}+\Omega_{\rm{r}}(1+z)^{4}+E^{\alpha}(z)(1-\Omega_{\rm{m}}-\Omega_{\rm{r}}).
\end{equation}
Obviously, the model with $\alpha=1$ reduces to the DGP model \cite{Dvali:2000hr} and with $\alpha=0$ reduces to the $\Lambda$CDM model.

\item GCG model: This model has an exotic EoS, $p_{\rm gcg}=-A/\rho_{\rm gcg}^\beta$, where $A$ is a positive constant and $\beta$ is a free parameter. The GCG behaves as a dust-like matter at the early times and behaves like a cosmological constant at the late times. In this model, we have
\begin{equation}
E^2(z)=\Omega_{\rm{b}}(1+z)^{3}+\Omega_{\rm{r}}(1+z)^{4}+(1-\Omega_{\rm{b}}-\Omega_{\rm{r}})\left(A_{\rm{s}}+(1-A_{\rm{s}})(1+z)^{3(1+\beta)}\right)^{1\over 1+\beta},
\end{equation}
where $A_{\rm{s}}\equiv A/\rho^{1+\beta}_{\rm{gcg}0}$ is a dimensionless parameter, $\rho_{\rm{gcg}0}$ is the present-day density of GCG, and $\Omega_{\rm b}$ is the present-day density of baryon matter. Actually, the GCG model can be viewed as a model of vacuum energy interacting with cold dark matter with the interaction term $Q=3\beta H \rho_\Lambda \rho_{\rm c}/(\rho_\Lambda+\rho_{\rm c})$, where $\rho_\Lambda$ and $\rho_{\rm c}$ are the energy densities of vacuum energy and cold dark matter. Thus, the GCG model with $\beta=0$ reduces to the $\Lambda$CDM model and with $\beta=1$ reduces to the original Chaplygin gas model.

\end{itemize}

\subsection{Current observations for cosmic distances}

\subsubsection{SN data}

We use the ``joint light-curve analysis'' (JLA) compilation \cite{Betoule:2014frx} of type Ia supernovae in this work. It consists of $740$ type Ia supernovae data, which are obtained by SDSS-II and SNLS collaborations. The peak luminosity of SN observation is correlated to stretch and color, so the distance modulus could be written as
\begin{equation}
\mu=m^{\ast}_{\rm{B}}-(M_{\rm{B}}-\alpha \times X_{1}+\beta \times C),
\end{equation}
where $m^{\ast}_{\rm{B}}$ is the peak magnitude of observation in a rest-frame B-band, $M_{\rm{B}}$ is absolute magnitude of a SN Ia, $X_{1}$ and $C$ are the time stretching of light curve and the supernova color at maximum brightness, respectively, and coefficients $\alpha$ and $\beta$ are the linear corrections, respectively, for stretch and color~\cite{Betoule:2014frx}. The luminosity distance $d_{\rm{L}}$ of a supernova can be calculated by
\begin{equation}
d_{{\rm L}}(z)=\frac{1+z}{H_{0}} \int_{0}^{z} \frac{dz'}{E(z')},
\end{equation}
where $E(z)=H(z)/H_0$ with $H_0$ being the Hubble constant. The $\chi^{2}$ function for SN observation is written as
\begin{equation}
\chi^{2}_{\rm{SN}}=(\hat{\mu}-\mu_{\rm{th}})^{\dagger}C_{\rm SN}^{-1}(\hat{\mu}-\mu_{\rm{th}}),
\end{equation}
where $C_{\rm SN}$ is the covariance matrix of the SN observation \cite{Betoule:2014frx}, and the theoretical distance modulus $\mu_{\rm{th}}$ can be calculated by                                                                                                \begin{equation}
\mu_{\rm{th}}=5\log_{10}\frac{d_{\rm{L}}}{10\rm{pc}}.
\end{equation}

\subsubsection{CMB data}

For the CMB data, we utilize the ``Planck distance priors'' from the Planck 2015 data~\cite{Ade:2015rim}. The shift parameter $R$ and the ``acoustic scale'' $\ell_{\rm{A}}$ can be given by
\begin{equation}
R\equiv\sqrt{\Omega_{\rm{m}}H^{2}_{0}}(1+z_{\ast})D_{\rm{A}}(z_{\ast}),
\end{equation}
\begin{equation}
\ell_{\rm{A}}\equiv(1+z_{\ast})\frac{\pi D_{\rm{A}}(z_{\ast})}{r_{\rm{s}}(z_{\ast})},\label{la}
\end{equation}
where $\Omega_{\rm{m}}$ denotes the present-day fractional energy density of matter, $z_{\ast}$ is the redshift at the decoupling epoch of photons, $D_{\rm{A}}(z_{\ast})$ denotes the angular diameter distance at the redshift $z_{\ast}$, and $r_{\rm s}(z_{\ast})$ denotes the comoving sound horizon at the redshift $z_{\ast}$. In a flat universe, $D_{\rm{A}}$ can be expressed as
\begin{equation}
D_{\rm{A}}(z)=\frac{1}{H_{0}(1+z)}\int_{0}^{z}\frac{dz'}{E(z')},\label{DA}
\end{equation}
and $r_{\rm{s}}(z)$ can be expressed as
\begin{equation}
r_{\rm{s}}(z)=\frac{1}{\sqrt{3}H_0}\int_{z}^{\infty}\frac{dz'}{(1+z')E(z')\sqrt{1+(3\Omega_{{\rm b}}/4\Omega_{{\rm \gamma}})(1+z')^{-1}}},
\label{rs}
\end{equation}
where $\Omega_{{\rm b}}$ and $\Omega_{{\rm \gamma}}$ are the present-day baryon energy density and the present-day photon energy density, respectively.
We take $3\Omega_{{\rm b}}/4\Omega_{{\rm \gamma}}=31500\Omega_{\rm{b}}h^{2}(T_{\rm{cmb}}/2.7{\rm K})^{-4}$, with $T_{\rm{cmb}}=2.7255$ K. $z_{\ast}$ is given by~\cite{Hu:1995en}
\begin{equation}
z_{\ast}=1048[1+0.00124(\Omega_{{\rm b}}h^{2})^{-0.738}][1+g_{1}(\Omega_{{\rm m}}h^{2})^{g_{2}}],
\end{equation}
where
\begin{equation}
g_{1}=\frac{0.0783(\Omega_{\rm{b}}h^{2})^{-0.238}}{1+39.5(\Omega_{\rm{b}}h^{2})^{-0.76}}, \;  g_{2}=\frac{0.560}{1+21.1(\Omega_{\rm{b}}h^{2})^{1.81}}.
\end{equation}
From the Planck TT+LowP data, the three values of the distance priors can be obtained: $R=1.7488\pm0.0074$, $\ell_{\rm{A}}=301.76\pm0.14$, and $\Omega_{\rm{b}}h^{2}=0.02228\pm0.00023$ \cite{Ade:2015rim}. The $\chi^{2}$ function for CMB is
\begin{equation}
\chi^{2}_{\rm{CMB}}=\Delta p_{i}[{\rm Cov}^{-1}_{\rm{CMB}}(p_{ i},p_{ j})]\Delta p_{ j},~\Delta p_{ i}=p_{ i}^{\rm{obs}}-p_{ i}^{\rm{th}},
\end{equation}
where $p_{1}=\ell_{\rm{A}}$, $p_{2}=R$, $p_{3}=\Omega_{\rm{b}}h^{2}$, and ${\rm Cov}^{-1}_{\rm CMB}$ is the inverse covariance matrix that can be found in Ref.~\cite{Ade:2015rim}.

\subsubsection{BAO data}

\begin{table*}[htbp]
\caption{Values of the distance ratio $\xi(z_{\rm eff})=r_{\rm d}/D_{\rm{V}}(z_{\rm eff})$ or $D_{\rm{V}}(z_{\rm eff})/r_{\rm d}$ from the BAO measurements. }
\label{tablebao}
\small
\setlength\tabcolsep{3.5pt}
\renewcommand{\arraystretch}{1.5}
\centering
\begin{tabular}{cccc}
\\
\hline\hline
$z_{\rm eff}$ & $\xi(z_{\rm eff})$ & Experiment & Reference\\
\hline
0.106 & $r_{\rm d}/D_{\rm{V}}(z_{\rm eff})=0.336\pm0.015$ & 6dFGS &~\cite{Beutler:2011hx} \\
0.15 &$D_{\rm{V}}(z_{\rm eff})/r_{\rm d}=4.466\pm0.168$ &SDSS-DR7 &~\cite{Ross:2014qpa}\\
0.32 & $D_{\rm{V}}(z_{\rm eff})/r_{\rm d}=8.467\pm0.167$ &BOSS-DR11 &~\cite{Anderson:2013zyy}\\
0.57 & $D_{\rm{V}}(z_{\rm eff})/r_{\rm d}=13.773\pm0.134$ &BOSS-DR11 &~\cite{Anderson:2013zyy} \\
\hline\hline
\end{tabular}
\end{table*}

Using the BAO measurements, we can obtain the ratio of the effective distance measure $D_{{\rm V}}(z)$ and the comoving sound horizon size $r_{\rm{s}}(z_{\rm{d}})$. The spherical average gives us the expression of $D_{\rm{V}}(z)$,
\begin{equation}
D_{\rm{V}}(z)\equiv\left[(1+z)^{2}D^{2}_{\rm{A}}(z)\frac{z}{H(z)}\right]^{1/3}.
\end{equation}
The comoving sound horizon size $r_{\rm s}(z_{\rm d})$ is given by Eq.~(\ref{rs}),
where $z_{\rm{d}}$ is the redshift of the drag epoch and its fitting formula is given by~\cite{Eisenstein:1997ik}
\begin{equation}
z_{\rm{d}}=\frac{1291(\Omega_{\rm{m}}h^2)^{0.251}}{1+0.659(\Omega_{\rm{m}}h^2)^{0.828}}[1+b_1(\Omega_{\rm{b}}h^2)^{b_2}],
\end{equation}
where
\begin{equation}
\begin{gathered}
b_1=0.313(\Omega_{\rm{m}}h^2)^{-0.419}[1+0.607(\Omega_{\rm{m}}h^2)^{0.674}],\\
b_2=0.238(\Omega_{\rm{m}}h^2)^{0.223}.
\end{gathered}
\end{equation}

We use four BAO data points: $r_{\rm d}/D_{\rm{V}}(z_{\rm eff})=0.336\pm0.015$ at $z_{\rm{eff}} = 0.106$ from the 6dF Galaxy Survey~\cite{Beutler:2011hx}, $D_{\rm{V}}(z_{\rm eff})(r_{\rm d,fid}/r_{\rm d})=664\pm25~{\rm Mpc}$ at $z_{\rm{eff}}=0.15$ from the SDSS-DR7 (where $r_{\rm d,fid} = 148.69~{\rm Mpc}$) \cite{Ross:2014qpa}, $D_{\rm{V}}(z_{\rm eff})(r_{\rm d,fid}/r_{\rm d})=1264\pm25~{\rm Mpc}$ at $z_{\rm{eff}}=0.32$, and $D_{{\rm V}}(z_{\rm eff})(r_{\rm d,fid}/r_{\rm d})=2056\pm20~{\rm Mpc}$ at $z_{\rm{eff}}=0.57$ from the BOSS-DR11 (where $r_{\rm d,fid}=149.28~{\rm Mpc}$) \cite{Anderson:2013zyy}. Note that here $r_{\rm d}\equiv r_{\rm s}(z_{\rm d})$ and thus $r_{\rm d,fid}$ is the sound horizon size of the drag epoch in a fiducial cosmology. For convenience, we give the values of the distance ratio $r_{\rm d}/D_{\rm{V}}(z_{\rm eff})$ or $D_{\rm{V}}(z_{\rm eff})/r_{\rm d}$ from the BAO measurements in Table \ref{tablebao}. The $\chi^{2}$ function for BAO is

\begin{equation}
\chi^2_{\rm BAO}=\sum\limits_{ i=1}^4 \frac{(\xi^{\rm obs}_{ i}-\xi^{\rm th}_{ i})^2}{\sigma_{ i}^2},
\end{equation}
where $\xi_{\rm th}$ and $\xi_{\rm obs}$ are, respectively, the theoretically predicted value and the corresponding experimentally measured value for the $i$th point of the BAO observation, and $\sigma_{i}$ is the standard deviation of the $i$-th point.

\subsection{Hubble parameter observations}

\subsubsection{Current $H(z)$ data}

\begin{table*}[htbp]
\caption{The Hubble parameter values $H(z)$ from the measurements of differential age of galaxies. Here $H(z)$ and $\sigma_{H(z)}$ are in units of km~s$^{-1}$~Mpc$^{-1}$.}
\label{tablehz}
\small
\setlength\tabcolsep{3.5pt}
\renewcommand{\arraystretch}{1.3}
\centering
\begin{tabular}{cccccccccc}
\\
\hline\hline
$z$  & $H(z)$ & $\sigma_{H(z)}$ & Reference \\ \hline

$0.07$   & $69.0$   & $19.6$    &~\cite{H1}   \\
$0.1$    & $69.0$    & $12.0$     &~\cite{H2}   \\
$0.12$   & $68.6$   & $26.2$    &~\cite{H1}  \\
$0.17$   & $83.0$   & $8.0$    &~\cite{H2} \\
$0.179$   & $75.0$   & $4.0$    &  ~\cite{H3}  \\
$0.1993$   & $75.0$   & $5.0$    &  ~\cite{H3}  \\
$0.2$   & $72.9$   & $29.6$    & ~\cite{H1} \\
$0.27$   & $77.0$   & $14.0$    &~\cite{H2}  \\
$0.28$   & $88.8$   & $33.6$    &~\cite{H1}   \\
$0.3519$   & $83.0$   & $14.0$    & ~\cite{H3}  \\
$0.3802$   & $83.0$   & $13.5$    &~\cite{Moresco:2016mzx}  \\
$0.4$   & $95.0$   & $17.0$    & ~\cite{H2}  \\
$0.4004$   & $77.0$   & $10.2$    &~\cite{H3} \\
$0.4247$   & $87.1$   & $11.2$    &~\cite{H3} \\
$0.4497$   & $92.8$   & $7.8$    &~\cite{Moresco:2016mzx}   \\
$0.4783$   & $80.9$   & $9.0$    & ~\cite{Moresco:2016mzx}  \\
$0.48$   & $97.0$   & $60.0$    & ~\cite{H2}  \\
$0.5929$   & $104.0$   & $13.0$    & ~\cite{H3}  \\
$0.6797$   & $92.0$   & $8.0$    & ~\cite{H3}   \\
$0.7812$   & $105.0$   & $12.0$    &~\cite{H3}   \\
$0.8754$   & $125.0$   & $17.0$    &~\cite{H3}  \\
$0.88$   & $90.0$   & $40.0$    & ~\cite{H2} \\
$0.9$   & $117.0$   & $23.0$    &~\cite{H2}   \\
$1.037$   & $154.0$   & $20.0$    & ~\cite{H3}  \\
$1.3$   & $168.0$   & $17.0$    &~\cite{H2}  \\
$1.363$   & $160.0$   & $33.6$    &  ~\cite{Moresco:2015cya} \\
$1.43$   & $177.0$   & $18.0$    & ~\cite{H2}   \\
$1.53$   & $140.0$   & $14.0$    &  ~\cite{H2}  \\
$1.75$   & $202.0$   & $40.0$    &  ~\cite{H2} \\
$1.965$   & $186.5$   & $50.4$    & ~\cite{Moresco:2015cya}  \\
\hline\hline
\end{tabular}
\end{table*}

In recent years, enormous efforts have been made in the measurements of $H(z)$ data. By definition, the Hubble parameter can be expressed as
\begin{equation}\label{2}
    H(z)=\frac{\dot{a}}{a}=-\frac{1}{1+z} \frac{dz}{dt},
\end{equation}
where ${a}$ and $\dot{a}$ are the cosmic scale factor and its rate of change with the cosmic time $t$, respectively.

According to the second equal sign in Eq.~(\ref{2}), we can get the $H(z)$ data by measuring the differential age (DA) of galaxies, which is called the DA method or the cosmic chromometer method. The method was proposed by Jimenez and Loeb in $2002$ \cite{Jimenez:2001gg} and then the first $H(z)$ measurement at $z\approx1$ was made in 2003~\cite{Jimenez:2003iv}.

We adopt $30$ data points obtained from the DA method in this work. Concretely, Stern et al. \cite{Stern} provided $11$ $H(z)$ data points covering the redshift range $0.1<z<1.75$ in $2010$ and discussed the role of the direct $H(z)$ measurements in constraining dark energy parameters by combining the $H(z)$ data with CMB data. In $2012$, Moresco et al. \cite{Moresco:2012} reported $8$ $H(z)$ data points covering the redshift range $0.15<z<1.1$ obtained via the differential spectroscopic evolution of early-type galaxies. In $2014$, Zhang et al. \cite{Zhang} obtained $4$ data points from luminous red galaxies (LRGs) of Sloan Digital Sky Survey Data Release Seven (SDSS DR7) at the redshift range of $0<z<0.4$, and they also combined these four data with previous $H(z)$ data to constrain both the flat and non-flat $\Lambda$CDM models. About one years later, Moresco \cite{Moresco:2015cya} presented $2$ data points on the basis of previous dataset, and utilized these data to estimate the improvement of accuracy on cosmological parameters in the $\Lambda$CDM model and the $w$CDM model, finding an about $5\%$ improvement of $\Omega_{\rm m}$ and $w_{\rm 0}$. In addition, in $2016$ Moresco et al. \cite{Moresco:2016mzx} enriched the dataset again with five new model-independent $H(z)$ measurements around the redshift $z\sim0.45$ (in a narrow range) by using the cosmic chronometer approach. For convenience, we list these $H(z)$ data in Table \ref{tablehz}.

In the current literature~\cite{Duan:2016zdv,Farooq:2013hq,Chuang:2012hq,Chen:2011hq,Guo:2015gpa,Magana:2017nfs}, these $30$ data points have been used to estimate the accuracies of cosmological parameters in the $\Lambda$CDM and $w$CDM models, which provide constraints on dark energy comparable to or better than those provided by SN data. Note here that in this paper we do not adopt the $H(z)$ data compilation from the measurement of clustering of galaxies or quasars (simply called the ``clustering'' method). The main reason is that some authors have questioned that these data are not totally model-independent and thus cannot be adopted in cosmological parameter constraints \cite{Liao:2012zza,Liao:2012gq,Melia:2013sxa,Chen:2014tdy,Li:2015nta,Melia:2015nwa,Cai:2015pia,Blake:2012hq,Bautista:2017zgn,Farooq:2016zwm}. In addition, in this paper we also use the $4$ data from BAO observation, which is overlapping with the $H(z)$ data from the ``clustering'' method.

The $\chi^2$ function of $H(z)$ is given by
\begin{equation}\label{3}
   \chi^{2}_{{H(z)}}=\sum^{\rm N}_{i=1}\frac{[H^{\rm th}(z_{i})-H^{\rm obs}(z_{ i})]^{2}}{\sigma^{2}_{H(z_i)}},
\end{equation}
where $H^{\rm{th}}$ and $H^{\rm{obs}}$ are the theoretical value in a cosmological model and the measured value of $H(z)$, respectively, and $\sigma_{H(z_i)}$ is the standard deviation of the $i$-th data point.

\subsubsection{Future $H(z)$ data from SL test}

We also use the future SL test observation (simulated data). The method was proposed by Sandage in 1962 in order to directly probe the dynamics of the expansion \cite{sandage}. Furthermore, it was improved in 1998 by Loeb who suggested a possible scheme by decades-long observation of the redshift variation of distant quasars (QSOs) Lyman-$\alpha$ absorption lines \cite{Loeb:1998bu}. With the continuous development of spectroscopy in the near future, the CODEX experiment (COsmic Dynamics EXperiment) proposed for E-ELT will be capable of accessing the redshift range of $z\in[2,5]$~\cite{Liske:2008ph}. Thus, a new ``cosmological window'' will be opened by the SL test.

In order to generate the mock data of SL test, we adopt the scheme accordant with our previous papers~\cite{Geng:2018pxk,Guo:2015gpa,Geng:2014hoa,Geng:2014ypa,Geng:2015hen,He:2016rvp,Geng:2015ara}. The redshift variation, which is the main observable, is usually expressed as a spectroscopic velocity shift~\cite{Loeb:1998bu},
\begin{equation}\label{3}
   \Delta v=\frac{\Delta z}{1+z} =H_{\rm{0}} \Delta t_{\rm o} \left[1-\frac{E(z)}{1+z}\right],
\end{equation}
where $\Delta t_{\rm o}$ is the observer's time interval, and $E(z)=H(z)/H_{0}$ can be determined by specific cosmological models. According to Monte Carlo simulations carried out to Lyman-$\alpha$ absorption lines from CODEX, the standard deviation on $\Delta v$ can be estimated as~\cite{Liske:2008ph}
\begin{equation}\label{4}
   \sigma_{\Delta {v}} = 1.35\left(\frac{2370}{S/N}\right)\left(\frac{N_{\rm QSO}}{30}\right)^{-1/2}\left(\frac{1+z_{\rm QSO}}{5}\right)^{x} ~\mathrm{cm}~\mathrm{s}^{-1},
\end{equation}
where $x$ is $1.7$ for $2\leq z\leq4$ and $0.9$ for $z\geq4$, $S/N=3000$ is spectral signal-to-noise ratio of Lyman-$\alpha$, and $N_{\rm QSO}$ and $z_{\rm QSO}$ are the number and redshift of observed quasars, respectively.
In this work, we simulate $30$ $H(z)$ data within the redshift range of $z\in[2,5]$ in a $10$-year observation of redshift drift. These mock SL test data are uniformly distributed over six redshift bins of $z_{\rm QSO}\in[2,5]$. The fiducial cosmology for the SL simulated data is chosen to be the best fit result of the analysis of the data combination of SN+CMB+BAO+$H(z)$.

\section{Results and discussion}\label{sec:Results and Discussions}

\begin{table*}[htbp]
\caption{Fit results of the $\Lambda$CDM, $w$CDM, $\alpha$DE, and GCG models by using the SN+CMB+BAO, SN+CMB+BAO+$H(z)$, and SN+CMB+BAO+$H(z)$+SL data.}
\label{table2}
\small
\setlength\tabcolsep{2.8pt}
\renewcommand{\arraystretch}{1.5}
\centering
\begin{tabular}{cccccccccccc}
\\
\hline\hline &\multicolumn{2}{c}{SN+CMB+BAO}&& \multicolumn{2}{c}{SN+CMB+BAO+$H(z)$}&& \multicolumn{2}{c}{SN+CMB+BAO+$H(z)$+SL} \\
 \cline{2-3}\cline{5-6}\cline{8-9}
Parameter  & $\Lambda$CDM & $w$CDM && $\Lambda$CDM & $w$CDM && $\Lambda$CDM & $w$CDM \\ \hline

$w$        & $-1~({\rm fixed})$
                   &  $-0.964^{+0.034}_{-0.035}$&
                   & $-1~({\rm fixed})$
                   & $ -0.969_{-0.031}^{+0.036}$
                   && $-1~({\rm fixed})$
                   & $-0.965_{-0.034}^{+0.032}$\\

$\Omega_{\rm m}$
                  & $0.322^{+0.008}_{-0.007}$
                   & $0.327_{-0.009}^{+0.008}$&
                   & $0.322_{-0.008}^{+0.008}$
                   & $0.325_{-0.008}^{+0.009}$
                  & & $0.322^{+0.005}_{-0.004}$
                   & $0.325_{-0.005}^{+0.005}$\\

$h$
                   & $0.668_{-0.006}^{+0.005}$
                   &  $0.662^{+0.008}_{-0.008}$&
                   & $0.669_{-0.005}^{+0.006}$
                   &  $0.664_{-0.008}^{+0.007}$
                   && $0.669_{-0.004}^{+0.004}$
                   & $0.663^{+0.006}_{-0.006}$\\
               \cline{2-3}\cline{5-6}\cline{8-9}
Parameter  & $\alpha$DE & GCG & & $\alpha$DE & GCG   & &$\alpha$DE & GCG \\
\hline
$\alpha$
                   & $0.127^{+0.128}_{-0.125}$
                   & $-$&
                   & $ 0.133_{-0.112}^{+0.138}$
                   & $-$
                   && $0.118_{-0.122}^{+0.116}$
                   & $-$\\
$\beta$
                    & $-$
                   & $-0.045^{+0.063}_{-0.065}$&
                   & $-$
                   & $-0.040^{+0.060}_{-0.065}$
                  & & $-$
                   & $-0.047^{+0.027}_{-0.026}$\\
$A_{\rm s}$
                    & $-$
                   & $0.690^{+0.023}_{-0.028}$&
                   & $-$
                   & $0.693^{+0.023}_{-0.027}$
                   && $-$
                   & $0.690^{+0.006}_{-0.006}$\\

$\Omega_{\rm m}$   & $0.326^{+0.009}_{-0.008}$
                   & $0.346^{+0.027}_{-0.023}$&
                   & $0.325_{-0.008}^{+0.008}$
                   & $0.343^{+0.026}_{-0.022}$
                   && $0.324^{+0.005}_{-0.005}$
                   & $0.345^{+0.006}_{-0.006}$\\

$h$
                    & $0.662_{-0.008}^{+0.008}$
                   & $0.661_{-0.008}^{+0.008}$&
                   & $0.662_{-0.007}^{+0.009}$
                   & $0.662_{-0.007}^{+0.008}$
                   && $0.664_{-0.005}^{+0.005}$
                   & $0.661_{-0.004}^{+0.004}$\\
\hline\hline
\end{tabular}
\end{table*}

\begin{table*}[htbp]
\caption{Constraint errors of cosmological parameters in the $\Lambda$CDM, $w$CDM, $\alpha$DE, and GCG models by using the SN+CMB+BAO, SN+CMB+BAO+$H(z)$, and SN+CMB+BAO+$H(z)$+SL data.}
\label{table3}
\small
\setlength\tabcolsep{2.8pt}
\renewcommand{\arraystretch}{1.5}
\centering
\begin{tabular}{cccccccccccc}
\\
\hline\hline &\multicolumn{2}{c}{SN+CMB+BAO}&& \multicolumn{2}{c}{SN+CMB+BAO+$H(z)$}&& \multicolumn{2}{c}{SN+CMB+BAO+$H(z)$+SL} \\
 \cline{2-3}\cline{5-6}\cline{8-9}
Error  & $\Lambda$CDM & $w$CDM && $\Lambda$CDM & $w$CDM && $\Lambda$CDM & $w$CDM \\ \hline

$\sigma(w)$        & $-$
                   & $0.035$&
                   & $-$
                   & $0.034$
                   && $-$
                   & $0.033$\\

$\sigma(\Omega_{\rm m})$
                  & $0.008$
                   & $0.009$&
                   & $0.008$
                   & $0.009$
                  & & $0.005$
                   & $0.005$\\

$\sigma(\emph{h})$
                   & $0.006$
                   & $0.008$&
                   & $0.006$
                   & $0.008$
                   && $0.004$
                   & $0.006$\\
               \cline{2-3}\cline{5-6}\cline{8-9}
 Error  & $\alpha$DE & GCG & & $\alpha$DE & GCG   & &$\alpha$DE & GCG \\
\hline
$\sigma(\alpha)$
                   & $0.127$
                   & $-$&
                   & $0.126$
                   & $-$
                   && $0.119$
                   & $-$\\
$\sigma(\beta)$
                    & $-$
                   & $0.064$&
                   & $-$
                   & $0.062$
                  & & $-$
                   & $0.027$\\
$\sigma(A_{\rm s})$
                    & $-$
                   & $0.026$&
                   & $-$
                   & $0.025$
                   && $-$
                   & $0.006$\\

$\sigma(\Omega_{\rm m})$
                   & $0.009$
                   & $0.025$&
                   & $0.008$
                   & $0.024$
                   && $0.005$
                   & $0.006$\\

$\sigma(\emph{h})$
                    & $0.008$
                   & $0.008$&
                   & $0.008$
                   & $0.008$
                   && $0.005$
                   & $0.004$\\
\hline\hline
\end{tabular}
\end{table*}

\begin{table*}[htbp]
\caption{Constraint precisions of cosmological parameters in the $\Lambda$CDM, $w$CDM, $\alpha$DE, and GCG models by using the SN+CMB+BAO, SN+CMB+BAO+$H(z)$, and SN+CMB+BAO+$H(z)$+SL data.}
\label{table4}
\small
\setlength\tabcolsep{2.8pt}
\renewcommand{\arraystretch}{1.5}
\centering
\begin{tabular}{cccccccccccc}
\\
\hline\hline &\multicolumn{2}{c}{SN+CMB+BAO}&& \multicolumn{2}{c}{SN+CMB+BAO+$H(z)$}&& \multicolumn{2}{c}{SN+CMB+BAO+$H(z)$+SL} \\
 \cline{2-3}\cline{5-6}\cline{8-9}
Precision  & $\Lambda$CDM & $w$CDM && $\Lambda$CDM & $w$CDM && $\Lambda$CDM & $w$CDM \\
\hline
$\varepsilon(w)$        & $-$
                   & $0.036$&
                    & $-$
                   & $0.035$&
                   & $-$
                   & $0.034$\\

$\varepsilon(\Omega_{\rm m})$
                  & $0.025$
                   & $0.028$&
                   & $0.025$
                   & $0.028$
                  & & $0.016$
                   & $0.015$\\

$\varepsilon(\emph{h})$
                   & $0.009$
                   & $0.012$&
                   & $0.009$
                   & $0.012$
                   && $0.006$
                   & $0.009$\\
               \cline{2-3}\cline{5-6}\cline{8-9}
Precision & $\alpha$DE & GCG & & $\alpha$DE & GCG   & &$\alpha$DE & GCG \\
\hline
$\varepsilon(\alpha)$
                   & $1.000$
                   & $-$&
                    & $0.947$
                    &$-$&
                    &$1.008$
                   & $-$&\\
$\varepsilon(\beta)$
                    & $-$
                   & $1.422$&
                   & $-$
                   & $1.550$
                  && $-$
                   & $0.574$\\
$\varepsilon(A_{\rm s})$
                      & $-$
                   & $0.038$&
                   & $-$
                   & $0.036$
                  && $-$
                   & $0.009$\\

$\varepsilon(\Omega_{\rm m})$
                     & $0.028$
                   & $0.072$&
                   & $0.025$
                   & $0.070$
                   && $0.015$
                   & $0.017$\\

$\varepsilon(\emph{h})$
                  & $0.012$
                   & $0.012$&
                   & $0.012$
                   & $0.012$
                   && $0.008$
                   & $0.006$\\
\hline\hline
\end{tabular}
\end{table*}

\begin{figure*}[htbp]
\centerline{\includegraphics[width=16cm]{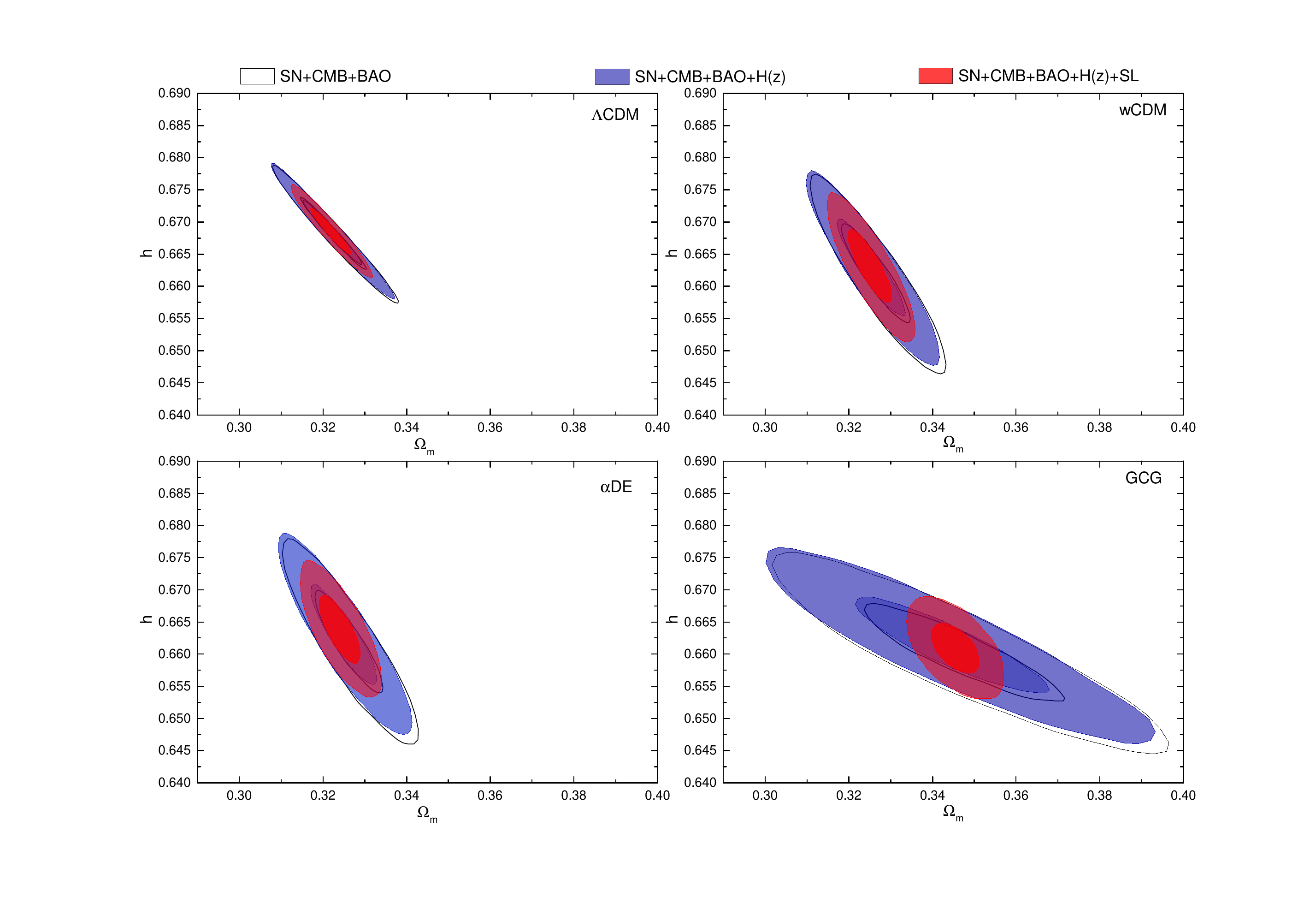}}
\caption{\label{fig2} Constraints ($1\sigma$ and 2$\sigma$ CL) on the $\Lambda$CDM, $w$CDM, $\alpha$DE, and GCG models in the $\Omega_{\rm m}$--$h$ plane by using the SN+CMB+BAO, SN+CMB+BAO+$H(z)$, and SN+CMB+BAO+$H(z)$+SL data.}
\end{figure*}

\begin{figure*}[htbp]
\centerline{\includegraphics[width=16cm]{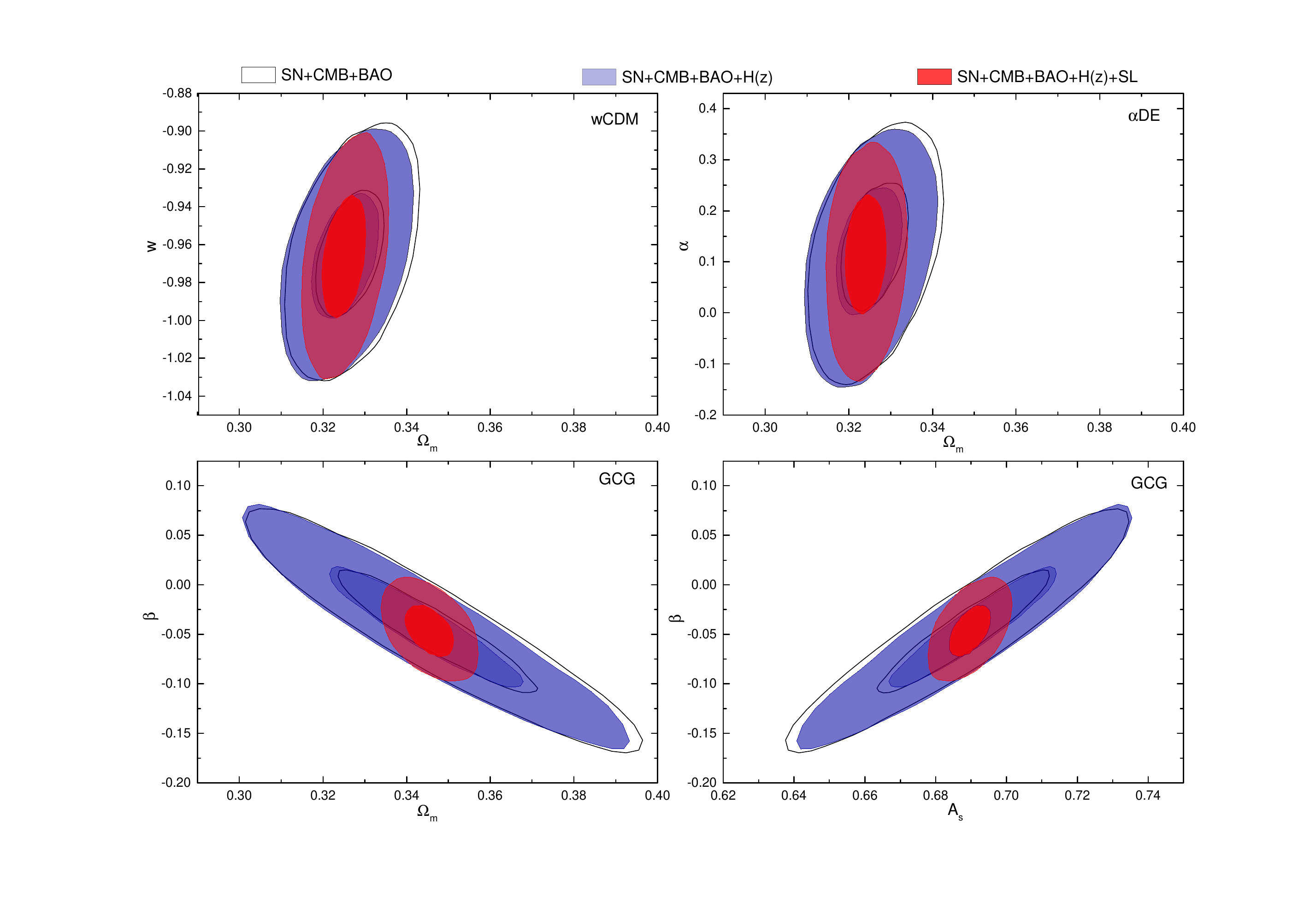}}
\caption{\label{fig3} Constraints ($1\sigma$ and 2$\sigma$ CL) on the $w$CDM, $\alpha$DE, and GCG models  from the SN+CMB+BAO, SN+CMB+BAO+$H(z)$, and SN+CMB+BAO+$H(z)$+SL data. We show the two-dimensional marginalized contours in the ${\Omega_{\rm{m}}}-{w}$ plane for $w$CDM, in the ${\Omega_{\rm{m}}}-{\alpha}$ plane for $\alpha$DE, in the ${\Omega_{\rm{m}}}-{\beta}$ and ${A_{\rm{s}}}-{\beta}$ planes for GCG.}
\end{figure*}


We constrain the $\Lambda$CDM, $w$CDM, $\alpha$DE, and GCG models by using the data combinations of SN+CMB+BAO, SN+CMB+BAO+$H(z)$, and SN+CMB+BAO+$H(z)$+SL. The detailed fit results are presented in Table~\ref{table2} with the $¡À1\sigma$ errors quoted. From Table~\ref{table2}, we find that, the constraint results from the SN+CMB+BAO+$H(z)$ data combination are very similar with those from the SN+CMB+BAO data combination, indicating that the current $H(z)$ data could not help improve the constraints on dark energy on the basis of the current cosmological observations (measuring the distance-redshift relation). But when the 10-year SL mock data are added to the SN+CMB+BAO+$H(z)$ data combination, the constraint results are significantly improved over other two cases, showing that the future $H(z)$ data from the SL test will have important potential to change the status of the Hubble parameter measurements in constraining dark energy.

In order to visually see the improvements of the parameter constraints with the addition of the current $H(z)$ data and the 10-year SL mock data, respectively, we present the constraints ($1\sigma$ and 2$\sigma$ CL) on the $\Lambda$CDM, $w$CDM, $\alpha$DE, and GCG models by using the SN+CMB+BAO, SN+CMB+BAO+$H(z)$, and SN+CMB+BAO+$H(z)$+SL data in Figs.~\ref{fig2} and~\ref{fig3}. From the two figures, we clearly see that adding the current $H(z)$ data to the SN+CMB+BAO data combination could not improve the constraints on the four dark energy models, and further adding the 10-year SL mock data to the SN+CMB+BAO+$H(z)$ data combination can lead to a tremendous improvement for the constraints on these models, in particular for the GCG model. In addition, we find that these mock $H(z)$ data from SL test can effectively break the degeneracies between ${\Omega_{\rm{m}}}$ and other parameters, in particular between ${\Omega_{\rm{m}}}$ and $\beta$ for the GCG model. Note that here the constraints from the 10-year SL mock data alone are not shown, because the SL data alone can only give rather weak constraints on dark energy models. Although the constraints from the SL data alone are very weak, they are still rather helpful in breaking the parameter degeneracies since the degeneracy directions formed by them in the parameter planes are different from those formed by the SN+CMB+BAO+$H(z)$ data. For the relevant issue, we refer the reader to see Fig.~2 of Ref.~\cite{Geng:2014hoa} and Fig.~4 of Ref.~\cite{Geng:2014ypa}, where the constraints on the $\Lambda$CDM and $w$CDM models from the 30-year SL mock data alone are shown.

With the purpose of quantifying the constraint improvements from the current $H(z)$ data and the 10-year SL mock data, we list the constraint errors for the $\Lambda$CDM, $w$CDM, $\alpha$DE, and GCG models by using the SN+CMB+BAO, SN+CMB+BAO+$H(z)$, and SN+CMB+BAO+\\$H(z)$+SL data in Table~\ref{table3}. We calculate the error as $\sigma=[({\sigma^2_+}+{\sigma^2_-})/2]^{1/2}$, where $\sigma_+$ and $\sigma_-$ correspond to the 1$\sigma$ deviation for upper and lower limits, respectively. Furthermore, for a parameter $\xi$, its constraint precision $\varepsilon(\xi)$ is defined as $\varepsilon(\xi)=\sigma(\xi)/{\xi_{\rm{bf}}}$, where $\xi_{\rm{bf}}$ is the best-fit value of $\xi$. In Table~\ref{table4}, we also present the constraint precisions of parameters of the four models by using the SN+CMB+BAO, SN+CMB+BAO+$H(z)$, and SN+CMB+BAO+$H(z)$+SL data.

From Table~\ref{table3}, we find that, with the addition of the current $H(z)$ data, the constraint errors of parameters are extremely close to those calculated from the SN+CMB+BAO data (without the current $H(z)$ data). But, when the 10-year SL mock data are combined to the SN+CMB+BAO+$H(z)$ data, the constraint errors of these parameters are greatly reduced. In other words, the current $H(z)$ data do not improve the constraints, but the 10-year SL mock data can obviously improve the constraints on dark energy in the four models.

Similarly, we show the constraint precisions of parameters for the $\Lambda$CDM, $w$CDM, $\alpha$DE, GCG models in Table~\ref{table4}. We see that, with the addition of the current $H(z)$ data, slight improvement is generated for the constraint precisions of parameters in the four models. When further adding the 10-year SL mock data to the SN+CMB+BAO+$H(z)$ data, the constraint precisions of parameters are tremendously improved for the four dark energy models. Concretely, the precision of ${\Omega_{\rm{m}}}$ is improved from 2.5$\%$ to 1.6$\%$ for $\Lambda$CDM, from 2.8$\%$ to 1.5$\%$ for $w$CDM, from 2.5$\%$ to 1.5$\%$ for $\alpha$DE, and from 7.0$\%$ to 1.7$\%$ for GCG. For the parameter $h$, the constraint precision is improved from 0.9$\%$ to 0.6$\%$ for $\Lambda$CDM, from 1.2$\%$ to 0.9$\%$ for $w$CDM, from 1.2$\%$ to 0.8$\%$ for $\alpha$DE, and from 1.2$\%$ to 0.6$\%$ for GCG.
In the $w$CDM model, the precision of $w$ is improved from 3.5$\%$ to 3.4$\%$; in the GCG model, the precision of $A_{\rm s}$ is improved from 3.6$\%$ to 0.9$\%$ and the precision of $\beta$ is improved from 155.0$\%$ to 57.4$\%$. For the parameter $\beta$ in the GCG model, its constraint result approaches 0, indicating that the $\Lambda$CDM limit of the GCG model is more favored by the observational data ($\beta\sim -0.04$ from SN+CMB+BAO, SN+CMB+BAO+$H(z)$, and SN+CMB+BAO+$H(z)$+SL). Obviously, this is due to the fact that the fiducial cosmology for the SL simulated data is chosen to be the same of the best fit of the SN+BAO+CMB+H(z) analysis and therefore the SL mock data can only tighten constraints around this cosmology. Note that $\beta=0$ is the $\Lambda$CDM limit and $\beta=1$ is the pure Chaplygin gas limit of the GCG model, and $\beta$ is not necessarily a positive number. We see that for the GCG model the error of $\beta$ is reduced by 56.5\% and for the $\alpha$DE model the error of $\alpha$ is reduced by 5.55\% once the 10-year SL data are considered. Therefore, we conclude that the current $H(z)$ data do not help improve the constraints on dark energy on the basis of the SN+CMB+BAO combination, but the 10-year SL data would significantly improve the constraints on dark energy on the basis of the SN+CMB+BAO+$H(z)$ combination.

\section{Conclusion}\label{sec:Conclusion}

In this work, we wish to investigate whether the current $H(z)$ measurements could help improve the constraints on dark energy on the basis of the mainstream cosmological probes (namely SN+CMB+BAO). We consider 30 $H(z)$ data measured using the DA method. Furthermore, we also consider the future redshift drift measurements (i.e., the SL test) by means of the E-ELT that is still in construction. We thus simulate 30 future $H(z)$ data from the 10-year redshift-drift observation. Thus, the second aim of this work is to investigate how the SL test changes the status of the Hubble parameter measurements in constraining dark energy. To fulfill the task, we employ several concrete dark energy models, including the $\Lambda$CDM, $w$CDM, $\alpha$DE, and GCG models, which are still favored by the current observations, at least to some extent.

We first use the data combination of SN+CMB+BAO to constrain the four dark energy models, and then we consider the addition of the current 30 $H(z)$ data in the data combination, i.e., we use the data combination of SN+CMB+BAO+$H(z)$ to constrain the models. We find that the current $H(z)$ data actually could not help improve the constraints on dark energy on the basis of the current mainstream cosmological probes (SN+CMB+BAO). Furthermore, we consider 30 mock SL test data simulated according to a 10-year observation in the data combination. By using the data combination SN+CMB+BAO+$H(z)$+SL, we find that compared to the case of SN+CMB+BAO+$H(z)$, the constraints on cosmological parameters are tremendously improved for all the four dark energy models.
For example, with the help of the 10-year SL observation, the constraint on ${\Omega_{\rm{m}}}$ is improved by about 40\%--75\% and the constraint on $H_0$ is improved by about 25\%--50\%. We find that the degeneracies between cosmological parameters could be effectively broken by the 10-year SL test data. Therefore, we conclude that although the current  $H(z)$ measurements could not improve the constraints on dark energy on the basis of the mainstream cosmological probes, the future redshift-drift observation will provide a good chance to change the status of the Hubble parameter measurements in constraining dark energy.

\begin{acknowledgments}
This work was supported by the National Natural Science Foundation of China (Grant Nos.~11875102, 11835009, 11522540, and 11690021) and the Top-Notch Young Talents Program of China.

\end{acknowledgments}


\begin{thebibliography}{99}

\bibitem{Riess:1998cb}
  A.~G.~Riess {\it et al.} [Supernova Search Team],
  Astron.\ J.\  {\bf 116}, 1009 (1998)
  [astro-ph/9805201].

\bibitem{Perlmutter:1998np}
  S.~Perlmutter {\it et al.} [Supernova Cosmology Project Collaboration],
  Astrophys.\ J.\  {\bf 517}, 565 (1999)
  [astro-ph/9812133].

\bibitem{Spergel:2003cb}
  D.~N.~Spergel et al. [WMAP Collaboration],
  Astrophys.\ J.\ Suppl.\  {\bf 148}, 175 (2003)
  [astro-ph/0302209].

\bibitem{Bennett:2003bz}
  C.~L.~Bennett et al. [WMAP Collaboration]
  Astrophys.\ J.\ Suppl.\  {\bf 148}, 1 (2003)
  [astro-ph/0302207].

\bibitem{Tegmark:2003ud}
  M.~Tegmark et al. [SDSS Collaboration],
  Phys.\ Rev.\ D {\bf 69}, 103501 (2004)
  [astro-ph/0310723].

\bibitem{Abazajian:2004aja}
  K.~Abazajian et al. [SDSS Collaboration],
  Astron.\ J.\  {\bf 128}, 502 (2004)
  [astro-ph/0403325].

\bibitem{Sahni:1999gb}
  V.~Sahni and A.~A.~Starobinsky,
  Int.\ J.\ Mod.\ Phys.\ D {\bf 9}, 373 (2000)
  [astro-ph/9904398].

\bibitem{Padmanabhan:2002ji}
  T.~Padmanabhan,
  Phys.\ Rep.\  {\bf 380}, 235 (2003)
  [hep-th/0212290].

\bibitem{Peebles:2002gy}
  P.~J.~E.~Peebles and B.~Ratra,
  Rev.\ Mod.\ Phys.\  {\bf 75}, 559 (2003)
  [astro-ph/0207347].

\bibitem{Copeland:2006wr}
  E.~J.~Copeland, M.~Sami and S.~Tsujikawa,
  Int.\ J.\ Mod.\ Phys.\ D {\bf 15}, 1753 (2006)
  [hep-th/0603057].

\bibitem{Sahni:2006pa}
  V.~Sahni and A.~Starobinsky,
  Int.\ J.\ Mod.\ Phys.\ D {\bf 15}, 2105 (2006)
  [astro-ph/0610026].

\bibitem{Frieman:2008sn}
  J.~Frieman, M.~Turner and D.~Huterer,
  Ann.\ Rev.\ Astron.\ Astrophys.\  {\bf 46}, 385 (2008)
  [arXiv:0803.0982 [astro-ph]].

\bibitem{Li:2011sd}
  M.~Li, X.~D.~Li, S.~Wang and Y.~Wang,
  Commun.\ Theor.\ Phys.\  {\bf 56}, 525 (2011)
  [arXiv:1103.5870 [astro-ph.CO]].

\bibitem{Bamba:2012cp}
  K.~Bamba, S.~Capozziello, S.~Nojiri and S.~D.~Odintsov,
  Astrophys.\ Space Sci.\  {\bf 342}, 155 (2012)
  [arXiv:1205.3421 [gr-qc]].

\bibitem{Weinberg:2012es}
  D.~H.~Weinberg, M.~J.~Mortonson, D.~J.~Eisenstein, C.~Hirata, A.~G.~Riess and E.~Rozo,
  Phys.\ Rep.\  {\bf 530}, 87 (2013)
  [arXiv:1201.2434 [astro-ph.CO]].

\bibitem{Mortonson:2013zfa}
  M.~J.~Mortonson, D.~H.~Weinberg and M.~White,
  arXiv:1401.0046 [astro-ph.CO].

\bibitem{Ade:2015xua}
  P.~A.~R.~Ade {\it et al.} [Planck Collaboration],
  Astron.\ Astrophys.\  {\bf 594}, A13 (2016)
  [arXiv:1502.01589 [astro-ph.CO]].

\bibitem{Weinberg}
  S.~Weinberg,
  Rev.\ Mod.\ Phys.\ {\bf 61}, 1, (1989)

\bibitem{Zeldovich}
  Ya.~B.~Zeldovich,
  Sov.\ Phys.\ Uspeksi.\ {\bf 11}, 3, (1969)

\bibitem{Moresco:2012}
  M.~Moresco{\it et al.},
  JCAP {\bf 2012}, no. 08, 006(2012)
  [arXiv:1201.3609 [astro-ph.CO]]

\bibitem{Farooq:2012ju}
  O.~Farooq and B.~Ratra,
  Phys.\ Lett.\ B {\bf 723}, 1 (2013)
  [arXiv:1212.4264 [astro-ph.CO]].

\bibitem{Farooq:2013hq}
  O.~Farooq and B.~Ratra,
  Astrophys.\ J.\ {\bf 766}, L7 (2013)
  [arXiv:1301.5243 [astro-ph.CO]].

\bibitem{Farooq:2013eea}
  O.~Farooq, S.~Crandall and B.~Ratra,
  Phys.\ Lett.\ B {\bf 726}, 72 (2013)
  [arXiv:1305.1957 [astro-ph.CO]].

\bibitem{Chuang:2012hq}
  C.~H.~Chuang and Y.~Wang,
  Mon.\ Not.\ Roy.\ Astron.\ Soc.\  {\bf 435}, 255 (2013)
  [arXiv:1209.0210 [astro-ph.CO]].

\bibitem{Chen:2011hq}
  Y.~Chen and B.~Ratra,
  Phys.\ Lett.\ Roy.\ B.\  {\bf 703}, 406 (2011)
  [arXiv:1106.4294 [astro-ph.CO]].

\bibitem{Gaztanaga:2009hq}
  E.~Gazta$\tilde{\rm{n}}$aga , A.~Cabr$\acute{\rm{e}}$,
  Mon.\ Not.\ R.\ Astron.\ Soc.\  {\bf 399}, 1663 (2009)
  [arXiv:1106.4294 [astro-ph.CO]].

\bibitem{Geng:2018pxk}
  J.~J.~Geng, R.~Y.~Guo, A.~Wang, J.~F.~Zhang and X.~Zhang,
  Commun.\ Theor.\ Phys.\  {\bf 70}, no. 4, 445 (2018)
  [arXiv:1806.10735 [astro-ph.CO]].
\bibitem{Wang:2016wjr}
  Y.~Wang {\it et al.} [BOSS Collaboration],
  Mon.\ Not.\ Roy.\ Astron.\ Soc.\  {\bf 469}, no. 3, 3762 (2017)
  [arXiv:1607.03154 [astro-ph.CO]].

\bibitem{Guo:2015gpa}
  R.~Y.~Guo and X.~Zhang,
  Eur.\ Phys.\ J.\ C {\bf 76}, no. 3, 163 (2016)
  [arXiv:1512.07703 [astro-ph.CO]].

\bibitem{H1}
  C.~Zhang, H.~Zhang, S.~Yuan, T.~J.~Zhang and Y.~C.~Sun,
  Res.\ Astron.\ Astrophys.\  {\bf 14}, 1221 (2014)
  [arXiv:1207.4541 [astro-ph.CO]].

\bibitem{H2}
  D.~Stern, R.~Jimenez, L.~Verde, M.~Kamionkowski and S.~A.~Stanford,
  JCAP {\bf 1002}, 008 (2010)
  [arXiv:0907.3149 [astro-ph.CO]].

\bibitem{H3}
  M.~Moresco{\it et al.},
  JCAP {\bf 1208}, 006 (2012)
  [arXiv:1201.3609 [astro-ph.CO]].

\bibitem{H4}
  X.~Y.~Xu {\it et al.},
  Mon.\ Not.\ Roy.\ Astron.\ Soc.\ {\bf 431}, 2834 (2013)

\bibitem{H5}
  C.~Blake {\it et al.},
  Mon.\ Not.\ Roy.\ Astron.\ Soc.\ {\bf 425}, 405 (2012)

 \bibitem{H6}
  A.~L.~Ratsimbazafy {\it et al.},
  Mon.\ Not.\ Roy.\ Astron.\ Soc.\ {\bf 467}, 3239 (2017)

 \bibitem{Anderson:2013oza}
  L.~Anderson {\it et al.},
  Mon.\ Not.\ Roy.\ Astron.\ Soc.\  {\bf 439}, no. 1, 83 (2014)
  [arXiv:1303.4666 [astro-ph.CO]].

\bibitem{Font-Ribera:2013wce}
  A.~Font-Ribera {\it et al.} [BOSS Collaboration],
  JCAP {\bf 1405}, 027 (2014)
  [arXiv:1311.1767 [astro-ph.CO]].

\bibitem{Delubac:2014aqe}
  T.~Delubac {\it et al.} [BOSS Collaboration],
  Astron.\ Astrophys.\  {\bf 574}, A59 (2015)
  [arXiv:1404.1801 [astro-ph.CO]].

\bibitem{Simon:2004tf}
  J.~Simon, L.~Verde and R.~Jimenez,
  Phys.\ Rev.\ D {\bf 71}, 123001 (2005)
  [astro-ph/0412269].

\bibitem{Zhang1010}
  T.~J.~Zhang and C.~Ma,
  Adv.\ Astron.\ {\bf 2010}, 184284 (2010)
  [arXiv:1010.1307[astro-ph.CO]]

\bibitem{Farooq1211}
  O.~Farooq, D.~Mania and B.~Ratra,
  Astrophys.\ J.\ {\bf 764}, 138 (2013)
  [arXiv:1211.4253 [astro-ph.CO]].

\bibitem{Farooq1212}
 O.~Farooq and B.~Ratra,
 Phys.\ Lett.\ B  {\bf723}, 1 (2013)
 [arXiv:1212.4264 [astro-ph.CO]].

\bibitem{Melia:2013hsa}
  F.~Melia and R.~S.~Maier,
  Mon.\ Not.\ Roy.\ Astron.\ Soc.\  {\bf 432}, 2669 (2013)
  [arXiv:1304.1802 [astro-ph.CO]].

\bibitem{Li:2014yza}
  Y.~L.~Li, S.~Y.~Li, T.~J.~Zhang and T.~P.~Li,
  Astrophys.\ J.\  {\bf 789}, L15 (2014)
  [arXiv:1404.0773 [astro-ph.CO]].

\bibitem{Sahni:2014ooa}
  V.~Sahni, A.~Shafieloo and A.~A.~Starobinsky,
  Astrophys.\ J.\  {\bf 793}, no. 2, L40 (2014)
  [arXiv:1406.2209 [astro-ph.CO]].

\bibitem{Chen:2013vea}
  Y.~Chen, C.~Q.~Geng, S.~Cao, Y.~M.~Huang and Z.~H.~Zhu,
  JCAP {\bf 1502}, 010 (2015)
  [arXiv:1312.1443 [astro-ph.CO]].

\bibitem{Park:2018tgj}
  C.~G.~Park and B.~Ratra,
  arXiv:1809.03598 [astro-ph.CO].
\bibitem{Park:2017xbl}
  C.~G.~Park and B.~Ratra,
  arXiv:1801.00213 [astro-ph.CO].

\bibitem{Park:2018bwy}
  C.~G.~Park and B.~Ratra,
  arXiv:1803.05522 [astro-ph.CO].

\bibitem{Park:2018fxx}
  C.~G.~Park and B.~Ratra,
  Astrophys.\ J.\  {\bf 868}, no. 2, 83 (2018)
  [arXiv:1807.07421 [astro-ph.CO]].

\bibitem{sandage}
  A.~Sandage,
  Astrophys. J. {\bf 136}, 319 (1962).

\bibitem{Loeb:1998bu}
  A.~Loeb,
  Astrophys.\ J.\  {\bf 499}, L111 (1998)
  [astro-ph/9802122].



\bibitem{Klockner:2015rqa}
  H.~R.~Kl$\ddot{\rm o}$ckner {\it et al.},
  PoS AASKA {\bf 14}, 027 (2015)
  [arXiv:1501.03822 [astro-ph.CO]].

\bibitem{Martins:2016bbi}
  C.~J.~A.~P.~Martins, M.~Martinelli, E.~Calabrese and M.~P.~L.~P.~Ramos,
  Phys.\ Rev.\ D {\bf 94}, no. 4, 043001 (2016)
  [arXiv:1606.07261 [astro-ph.CO]].

\bibitem{Yu:2013bia}
  H.~R.~Yu, T.~J.~Zhang and U.~L.~Pen,
  Phys.\ Rev.\ Lett.\  {\bf 113}, 041303 (2014)
  [arXiv:1311.2363 [astro-ph.CO]].




\bibitem{Lazkoz:2017fvx}
  R.~Lazkoz, I.~Leanizbarrutia and V.~Salzano,
  Eur.\ Phys.\ J.\ C {\bf 78}, no. 1, 11 (2018)
  [arXiv:1712.07555 [astro-ph.CO]].

\bibitem{Corasaniti:2007bg}
  P.~S.~Corasaniti, D.~Huterer and A.~Melchiorri,
  Phys.\ Rev.\ D {\bf 75}, 062001 (2007)
  [astro-ph/0701433].

\bibitem{Geng:2014hoa}
  J.~J.~Geng, J.~F.~Zhang and X.~Zhang,
  JCAP {\bf 1407}, 006 (2014)
  [arXiv:1404.5407 [astro-ph.CO]].

 \bibitem{Geng:2014ypa}
  J.~J.~Geng, J.~F.~Zhang and X.~Zhang,
  JCAP {\bf 1412}, no. 12, 018 (2014)
  [arXiv:1407.7123 [astro-ph.CO]].
\bibitem{Geng:2015hen}
  J.~J.~Geng, R.~Y.~Guo, D.~Z.~He, J.~F.~Zhang and X.~Zhang,
  Front.\ Phys.\ (Beijing) {\bf 10}, 109501 (2015)
  [arXiv:1511.06957 [astro-ph.CO]].

\bibitem{Balbi:2007fx}
  A.~Balbi and C.~Quercellini,
  Mon.\ Not.\ R.\ Astron.\ Soc.\  {\bf 382}, 1623 (2007)
  [arXiv:0704.2350 [astro-ph]].

\bibitem{Zhang:2007zga}
  H.~B.~Zhang, W.~H.~Zhong, Z.~H.~Zhu and S.~He,
  Phys.\ Rev.\ D {\bf 76}, 123508 (2007)
  [arXiv:0705.4409 [astro-ph]].

\bibitem{Zhang:2010im}
  J.~F.~Zhang, L.~Zhang and X.~Zhang,
  Phys.\ Lett.\ B {\bf 691}, 11 (2010)
  [arXiv:1006.1738 [astro-ph.CO]].

\bibitem{Quercellini:2010zr}
  C.~Quercellini, L.~Amendola, A.~Balbi, P.~Cabella and M.~Quartin,
  Phys.\ Rep.\  {\bf 521}, 95 (2012)
  [arXiv:1011.2646 [astro-ph.CO]].

\bibitem{Martinelli:2012vq}
  M.~Martinelli, S.~Pandolfi, C.~J.~A.~P.~Martins and P.~E.~Vielzeuf,
  Phys.\ Rev.\ D {\bf 86}, 123001 (2012)
  [arXiv:1210.7166 [astro-ph.CO]].

\bibitem{Li:2013oba}
  Z.~Li, K.~Liao, P.~Wu, H.~Yu and Z.~H.~Zhu,
  Phys.\ Rev.\ D {\bf 88}, (2), 023003 (2013)
  [arXiv:1306.5932 [gr-qc]].

\bibitem{Zhang:2013zyn}
  M.~J.~Zhang and W.~B.~Liu,
  Eur.\ Phys.\ J.\ C {\bf 74}, 2863 (2014)
  [arXiv:1312.0224 [astro-ph.CO]].

\bibitem{sl5}
  S.~Yuan and T.~J.~Zhang,
  JCAP {\bf 1502}, 025 (2015)
  [arXiv:1311.1583 [astro-ph.CO]].
\bibitem{Liske:2008ph}
  J.~Liske et al.,
  Mon.\ Not.\ R.\ Astron.\ Soc.\  {\bf 386}, 1192 (2008)
  [arXiv:0802.1532 [astro-ph]].

\bibitem{He:2016rvp}
  D.~Z.~He, J.~F.~Zhang and X.~Zhang,
  Sci.\ China Phys.\ Mech.\ Astron.\  {\bf 60}, no. 3, 039511 (2017)
  [arXiv:1607.05643 [astro-ph.CO]].

\bibitem{Xu:2016grp}
  Y.~Y.~Xu and X.~Zhang,
  Eur.\ Phys.\ J.\ C {\bf 76}, no. 11, 588 (2016)
  [arXiv:1607.06262 [astro-ph.CO]].

\bibitem{Dvali:2003rk}
  G.~Dvali and M.~S.~Turner,
  astro-ph/0301510.

\bibitem{Dvali:2000hr}
  G.~R.~Dvali, G.~Gabadadze and M.~Porrati,
  Phys.\ Lett.\ B {\bf 485}, 208 (2000)
  [hep-th/0005016].

\bibitem{Betoule:2014frx}
  M.~Betoule et al. [SDSS Collaboration],
  Astron.\ Astrophys.\  {\bf 568}, A22 (2014)
  [arXiv:1401.4064 [astro-ph.CO]].

\bibitem{Ade:2015rim}
  P.~A.~R.~Ade {\it et al.} [Planck Collaboration],
  Astron.\ Astrophys.\  {\bf 594}, A14 (2016)
  [arXiv:1502.01590 [astro-ph.CO]].

\bibitem{Hu:1995en}
  W.~Hu and N.~Sugiyama,
  Astrophys.\ J.\  {\bf 471}, 542 (1996)
  [astro-ph/9510117].

\bibitem{Eisenstein:1997ik}
  D.~J.~Eisenstein and W.~Hu,
  Astrophys.\ J.\  {\bf 496}, 605 (1998)
  [astro-ph/9709112].

\bibitem{Beutler:2011hx}
  F.~Beutler {\it et al.},
  Mon.\ Not.\ Roy.\ Astron.\ Soc.\  {\bf 416}, 3017 (2011)
  [arXiv:1106.3366 [astro-ph.CO]].

\bibitem{Ross:2014qpa}
  A.~J.~Ross, L.~Samushia, C.~Howlett, W.~J.~Percival, A.~Burden and M.~Manera,
  Mon.\ Not.\ Roy.\ Astron.\ Soc.\  {\bf 449}, no.1, 835 (2015)
  [arXiv:1409.3242 [astro-ph.CO]].

\bibitem{Anderson:2013zyy}
  L.~Anderson {\it et al.} [BOSS Collaboration],
  Mon.\ Not.\ Roy.\ Astron.\ Soc.\  {\bf 441}, no. 1, 24 (2014)
  [arXiv:1312.4877 [astro-ph.CO]].

\bibitem{Jimenez:2001gg}
  R.~Jimenez and A.~Loeb,
  Astrophys.\ J.\  {\bf 573}, 37 (2002)
  [astro-ph/0106145].

\bibitem{Jimenez:2003iv}
  R.~Jimenez, L.~Verde, T.~Treu and D.~Stern,
  Astrophys.\ J.\  {\bf 593}, 622 (2003)
  [astro-ph/0302560].

\bibitem{Stern}
Stern, Daniel {\it et al.},
JCAP {\bf1002}, 008 (2010)
[arXiv:0907.3149 [astro-ph.CO]].

\bibitem{Zhang}
  C.~Zhang, H.~Zhang, S.~Yuan, T.~J.~Zhang, Y.~C.~Sun,
  Res.\ Astron.\ Astrophys.\ {\bf 14}, 1221 (2014)

\bibitem{Moresco:2015cya}
  M.~Moresco
  Mon.\ Not.\ Roy.\ Astron.\ Soc.\  {\bf 450}, no. 1, L16 (2015)
  [arXiv:1503.01116 [astro-ph.CO]].

\bibitem{Moresco:2016mzx}
  M.~Moresco {\it et al.},
  JCAP {\bf 1605}, no. 05, 014 (2016)
  [arXiv:1601.01701 [astro-ph.CO]].

\bibitem{Duan:2016zdv}
  X.~W.~Duan, M.~Zhou and T.~J.~Zhang,
  arXiv:1605.03947 [astro-ph.CO].

\bibitem{Magana:2017nfs}
  J.~Magana, M.~H.~Amante, M.~A.~Garcia-Aspeitia and V.~Motta,
  Mon.\ Not.\ Roy.\ Astron.\ Soc.\  {\bf 476}, 1036 (2018)
  [arXiv:1706.09848 [astro-ph.CO]].

\bibitem{Liao:2012zza}
  K.~Liao, S.~Cao, J.~Wang, X.~L.~Gong and Z.~H.~Zhu,
  Phys.\ Lett.\ B {\bf 710}, 17 (2012)
  [arXiv:1205.0972 [astro-ph.CO]].



\bibitem{Liao:2012gq}
  K.~Liao, Y.~Pan and Z.~H.~Zhu,
  Res.\ Astron.\ Astrophys.\  {\bf 13}, 159 (2013)
  [arXiv:1210.5021 [astro-ph.CO]].



\bibitem{Melia:2013sxa}
  F.~Melia,
  JCAP {\bf 1401}, 027 (2014)
  [arXiv:1312.5798 [astro-ph.CO]].

\bibitem{Chen:2014tdy}
  Y.~Chen, C.~Q.~Geng, C.~C.~Lee, L.~W.~Luo and Z.~H.~Zhu,
  Phys.\ Rev.\ D {\bf 91}, (4), 044019 (2015)
  [arXiv:1407.4303 [astro-ph.CO]].

\bibitem{Li:2015nta}
  Z.~Li, J.~E.~Gonzalez, H.~Yu, Z.~H.~Zhu and J.~S.~Alcaniz,
   Phys.\ Rev.\ D {\bf 93}, no. 4, 043014 (2016)
  [arXiv:1504.03269 [astro-ph.CO]].

\bibitem{Melia:2015nwa}
  F.~Melia and T.~M.~McClintock,
  Astron.\ J.\  {\bf 150}, 119 (2015)
  [arXiv:1507.08279 [astro-ph.CO]].

\bibitem{Cai:2015pia}
  R.~G.~Cai, Z.~K.~Guo and T.~Yang,
  Phys.\ Rev.\ D {\bf 93}, no. 4, 043517 (2016)
  [arXiv:1509.06283 [astro-ph.CO]].

\bibitem{Blake:2012hq}
  C.~Blake,
  Mon.\ Not.\ R.\ Astron.\ Soc.\  {\bf 425}, 405 (2012)
  [arXiv:1204.3674 [astro-ph.CO]].

\bibitem{Bautista:2017zgn}
  J.~E.~Bautista {\it et al.},
  Astron.\ Astrophys.\  {\bf 603}, A12 (2017)
  [arXiv:1702.00176 [astro-ph.CO]].


\bibitem{Farooq:2016zwm}
  O.~Farooq, F.~R.~Madiyar, S.~Crandall and B.~Ratra,
  Astrophys.\ J.\  {\bf 835}, no. 1, 26 (2017)
  [arXiv:1607.03537 [astro-ph.CO]].

\bibitem{Geng:2015ara}
  J.~J.~Geng, Y.~H.~Li, J.~F.~Zhang and X.~Zhang,
  Eur.\ Phys.\ J.\ C {\bf 75}, no. 8, 356 (2015)
  [arXiv:1501.03874 [astro-ph.CO]].



\end{thebibliography}
\end{document}